\documentstyle[prl,aps,twocolumn]{revtex}
\large  

\input epsf
\begin{document}
\draft \twocolumn[\hsize\textwidth
\columnwidth\hsize\csname
@twocolumnfalse\endcsname
\title{Superfluid Hydrodynamics of an Electron Gas in a Superstrong 
Magnetic Field}
\author{Eugene B. Kolomeisky} 
\address{Department of Physics, University of Virginia,
382 McCormick Road, P. O. Box 400714, Charlottesville, Virginia 22904-4714}
\maketitle
\begin{abstract}
We derive the equations of hydrodynamics of a fully polarized electron gas 
placed in a strong magnetic field.  These equations reveal the existence
of solitons -- immobile or propagating domain wall-like defects whose plane is 
perpendicular to the field direction.  The solitons are used to construct 
weakly excited stationary states, and novel nonuniform persistent current 
states of the system.    
\end{abstract}
\vspace{2mm} 
]

\narrowtext

Understanding the properties of matter subject to an extremely
strong magnetic field \cite{Ruder} has fundamental importance.  At
fields of order $10^{9} G$ the interaction of atomic electrons with
the external magnetic field begins to dominate over their Coulomb
interaction which results in a novel kind of atomic physics
\cite{Lieb}.  In semiconductors and semimetals due to the smallness of
the effective mass of the charge carriers and dielectric screening of
the Coulomb interaction
\cite{AM}, the corresponding field is substantially smaller, and
accessible in the laboratory \cite{Silin}.

Since a uniform magnetic field localizes the motion of a charged
particle in the plane perpendicular to the field direction, and does
not affect the motion along the field \cite{LL1}, strong fields can be
used to study essentially one-dimensional physics using
three-dimensional samples.  The one-dimensionality is most pronounced
if the field is strong enough to pull all the carriers into the lowest
Landau level.  In this limit in an interacting system a variety of
phases of one-dimensional nature such as charge-density waves
\cite{Silin,McB,Yak}, Wigner crystals \cite{McB}, and
excitonic insulators \cite{Abr} have been predicted, and a
uniform-density state (adiabatically evolving from the noninteracting
limit) occupies only a portion of the phase diagram
\cite{McB}. Recent work \cite{Maslov} has demonstrated that this phase
has Luttinger-liquid-like properties.

In this Letter, by studying elementary excitations, we show that the 
uniform-density phase behaves as a unidirectional superfluid.  The governing 
equations of hydrodynamics have a rich mathematical structure which 
owes its existence to the one-dimensional physics of the problem.  A detailed
treatment will be presented elsewhere \cite{KNSQ1}.  Below we restrict 
ourselves only to the salient physics effects.
   
We start with reminding the reader of the method of constructing the ground 
state of a noninteracting electron gas of density $n$ placed in a fully
polarizing magnetic field $H$ pointing along the $x$ direction of a 
coordinate system.  The motion of electrons is confined within
infinitely long Landau tubes parallel to the field direction: the
motion perpendicular to the tube axis is localized within the magnetic
length $l = (\hbar c/eH)^{1/2}$ while the motion along the tube is
free.  The degeneracy of the lowest Landau level, i.e. the number of Landau
tubes in the system of cross-sectional area $S$ perpendicular to the
field direction is given by \cite{LL1} $eHS/2\pi \hbar c = S/2\pi
l^{2}$ which implies that the cross-sectional area of the individual
tube is $2\pi l^{2}$.  Therefore the magnetic flux confined within the
tube is the flux quantum $hc/e$, and thus each Landau tube can be
thought of as a magnetic field line.  Hence the quantum state of
every electron can be characterized by specifying the Landau tube it
belongs to, and the wavevector $k_{x}$ describing the free motion
along the field.  Because of the Pauli principle, for a given Landau
tube all the states with $|k_{x}| \le k_{F}$ are occupied while all
the states having $|k_{x}| > k_{F}$ are empty.  The Fermi wave vector
$k_{F}$ can be found as $k_{F} = \pi \rho$ where $\rho$ is the linear
electron density in the tube, which can be computed by dividing the
total number of electrons $N$ by the number of Landau tubes $S/2\pi
l^{2}$, and by the tube length $L$.  Thus $\rho = 2\pi l^2n$, $k_{F} =
2\pi ^2 l^{2}n$ and the Fermi energy can be found as $\epsilon_{F} =
\hbar ^{2}k_{F}^{2}/2m = (2\pi ^2 \hbar l^{2}n)^{2}/2m$ where $m$ is 
the electron mass.  The expression for the ground-state energy
\begin{equation}
\label{grstate}
E_{0} = {V\over (2\pi l)^{2}} \int \limits _{-k_{F}}^{k_{F}}{\hbar
^{2}k_{x}^{2}\over 2m}dk_x = {V (2\pi ^2\hbar l^{2})^{2}n^{3} \over 6m}
\end{equation}
(where $V = SL$ is the system volume) can
be rewritten as $(L\pi ^{2}\hbar ^{2}/6m)(2\pi l^{2}n)^{3}$ times  
$S/2\pi l^{2}$: the ground-state energy of a one-dimensional spinless Fermi 
gas of linear density $\rho = 2\pi l^2n$, multiplied by the total number of 
Landau tubes. The condition of being at the lowest Landau
level is satisfied if the Fermi energy lies below the
bottom of the second Landau level $\hbar eH/mc$ \cite{LL1}.  This leads to 
the condition $nl^{3} \le 1/\pi ^{2}\sqrt{2}$
which gives the range of applicability of Eq.(\ref{grstate}).  In what
follows we will adopt the stronger condition $nl^{3} \ll 1$ which also allows
us to neglect transitions to the higher Landau levels.

The presence of the Coulomb interaction introduces a new length scale into
the problem, the ``exciton'' Bohr radius $a = \epsilon \hbar
^{2}/me^{2}$, where $\epsilon$ is the dielectric constant of the
medium.  In astrophysical applications when $\epsilon = 1$, and $m
= m_{e}$, the electron mass in vacuum, $a$ reduces to Bohr's radius.
In the limit $nl^{3} \ll 1$ the average distance between the electrons
belonging to the same Landau tube, $1/2\pi l^{2}n$, is much bigger than
the average distance between the nearest tubes $l$.  Thus the electrons 
belonging to the same
Landau tube interact much more weakly among themselves than the
electrons belonging to different tubes.  Therefore the physical
picture of independent Landau tubes survives if the Fermi energy
is much bigger than the Coulomb interaction of
electrons confined to different tubes.  The latter (in an electrically
neutral system) is estimated \cite{Silin,Onis} as the plasmon energy
$\hbar \omega_{p}$ where $\omega_{p}^{2} = 4\pi ne^{2}/m\epsilon_{0}$
is the plasma frequency.  Combining the requirement $\epsilon_{F} \gg \hbar
\omega_{p}$ with the condition that all electrons are at the bottom of
the lowest Landau level, we arrive at the concentration interval
\begin{equation}
\label{interval}
(a/l)^{8/3} \ll na^{3} \ll (a/l)^{3}
\end{equation}
which is the range of applicability of our theory.  The
interval (\ref{interval}) exists if the exciton Bohr radius $a$ is
much bigger than the magnetic length $l$; the magnetic fields
satisfying this condition are refered to as superstrong \cite{Silin}.
Both length scales coincide at the field $H_{0} =
m^{2}e^{3}c/\epsilon^{2}\hbar ^{3}$.  In astrophysical
applications ($m=m_{e}$, $\epsilon=1$) $H_{0} = 2.35 \cdot 10^{9}
G$, while in condensed matter applications the corresponding field is
$(\epsilon m_{e}/m)^{2} \sim 10^4 - 10^6$ times smaller, and for
$a/l
\sim 5 - 10$ the density range (\ref{interval}) exists in doped
semiconductors and semimetals.

We postulate that the long-wavelength physics of a polarized electron gas
placed in a slowly varying external potential $V({\bf r})$ is
described by the energy functional of the {\it complex} order parameter
$\Psi({\bf r},t)$:
\begin{equation}
\label{functional}      
F = \int dV \left[{\hbar^{2}\over 2m}\left|{d\Psi \over dx}\right|^{2} + V({\bf
r})|\Psi|^{2} + {(2\pi ^{2}\hbar l^{2})^{2} \over 6m}|\Psi|^{6}\right]
\end{equation}
The physical meaning of $\Psi$ is that $|\Psi|^2 = n({\bf r},t)$ gives
the electron density which is a function of position ${\bf r}$ and
time $t$.  If $\Psi$ is restricted to be real, then (\ref{functional})
can be recognized as a starting point of a generalized Thomas-Fermi
density functional theory \cite{Brack}: for the 
translationally-invariant case the density is uniform, and then
(\ref{functional}) reduces to the ground-state energy (\ref{grstate}).
The second term of (\ref{functional}) is the energy in the external
field. The derivative term, known as ``Weizsacker inhomogeneity 
correction'' \cite{Brack}, which is a lowest-order term of a gradient 
expansion, takes into account the fact that deviations from uniformity in the 
field direction are costly.  Density functionals similar to 
(\ref{functional}) are used to describe static properties of multielectron 
atoms placed in very strong magnetic fields \cite{Lieb}.

We further postulate that the order parameter $\Psi$ evolves according
to the equation of motion
\begin{equation}
\label{nlse}
i\hbar \partial_{t}\Psi = {\delta F \over \delta\Psi_{*}} = \left[-{\hbar
^{2} \partial ^{2}_{x} \over 2m} + V({\bf r}) + {(2\pi
^{2}\hbar l^{2})^{2}|\Psi|^{4} \over 2m}\right ]\Psi
\end{equation}
In the representation $\Psi = n^{1/2}\exp i\theta$ and $v = (\hbar
/m)\partial_{x}\theta$, Eq.(\ref{nlse}) reduces to the system of two
coupled equations:
\begin{equation}
\label{cont}
\partial_{t}n + \partial_{x}(nv) = 0                                           \end{equation}
\begin{equation}
\label{eiler}      
m(\partial_{t}v + v\partial_{x}v) = - \partial_{x}[V + {(2\pi
^2\hbar l^{2})^{2} \over 2m}n^2 - {\hbar ^{2} \partial
^{2}_{x}\sqrt{n}\over 2m \sqrt{n}}]
\end{equation}
In the long-wavelength limit when the last term in the right-hand side
of (\ref{eiler}) can be dropped, the system of equations (\ref{cont})
and (\ref{eiler}) can be identified with those of hydrodynamics of an
ideal fluid \cite {LL2} whose pressure and density are related by the
equation of state $p = (2\pi ^2 \hbar l^{2})^{2}n^{3}/3m$.  The
variable $v = (\hbar /m)\partial_{x}\theta$ has a meaning of the
superfluid velocity which only has a component along the
field direction.
We note that the equations of hydrodynamics of an electron gas in a
superstrong magnetic field could be written down from the outset and
the density dependence of pressure can be computed from the ground-state 
energy (\ref{grstate}) as $p = - dE/dV$.  This provides an important 
test of the underlying equations (\ref{functional})
and (\ref{nlse}); specifically, selection of the gradient term of
(\ref{functional}) in the Weizsacker form plays a central role in
getting the correct hydrodynamical limit. We stress that hydrodynamical
description requires two fields, density and velocity; that is why the
order parameter $\Psi$ must be complex. 

{\it Spectrum of small density oscillations:} - Let us look for
solutions to (\ref{nlse}) (with $V = 0$) of the form $\Psi(x,t) = \psi(x,t)
\exp (-i\mu t/\hbar)$.  The function $\psi(x,t)$ then obeys the equation
\begin{equation}
\label{uniform}
i\hbar \partial_{t}\psi = {\hbar ^{2} \over 2m}\left [-\partial ^{2}_{x} +
(2\pi ^2l^{2})^{2}(|\psi|^{4} - \psi _{0}^{4})\right ]\psi 
\end{equation}
where the chemical potential $\mu = \pi ^{2}\hbar ^{2}(2\pi
l^{2})^{2}\psi_{0}^{4}/2m$ is selected so that the ground state
corresponds to the uniform electron density $n_{0} = \psi_{0}^{2}$.
Consider a small oscillation of $\psi$ around its constant value $\psi
_{0}$: $\psi = \psi_{0} + A\exp [i(qx - \omega t)] + B^{*}\exp [-i(qx - \omega
t)]$, where $A$ and $B^{*}$ are small complex amplitudes, and $q$ is the
wavevector along the field direction.  Substituting the expression for
$\psi$ into (\ref{uniform}), linearizing, and solving the resulting system of two equations, we find for the dispersion law
\begin{equation}
\label{dispersion}
(\hbar \omega)^{2} = (\hbar ^{2}q^{2}/2m)^{2} + (\hbar cq)^{2}
\end{equation}                      
The spectrum takes the Bogoliubov form \cite{LP}: for large momenta
the dispersion law coincides with the free-particle energy $\hbar
^{2}q^{2}/2m$, while at low momenta we get instead the phonon
dispersion $\omega = cq$, where $c = 2\pi ^{2}l^{2}\hbar n_{0}/m$ is
the sound velocity.  We note that this velocity coincides with the
hydrodynamical expression $c = (dp/mdn)^{1/2}$ for a gas with equation
of state $p = (2\pi ^2 \hbar l^{2})^{2}n^{3}/3m$. The
Landau criterion applied to (\ref{dispersion}) then implies \cite{LP} that the
critical velocity of superfluid motion along the direction of magnetic
field equals the sound velocity.

{\it Solitons:} - Without the external potential ($V = 0$)
Eq.(\ref{uniform}) has soliton solutions \cite{KNSQ}.  These describe
moving (gray soliton) or standing (dark soliton) density depressions
accompanied by a corresponding profile of phase \cite{Kivshar}.  Because 
of the system
uniformity in the plane perpendicular to the magnetic field, the
solitons are domain wall-like defects whose plane is perpendicular to
the direction of magnetic field; the solitons can only propagate along
(or opposite) to the field direction (Figure 1).  
\begin{figure}[htbp]
\epsfxsize=3.0in
\vspace*{-0.3cm}
\hspace*{0.2cm}
\epsfbox{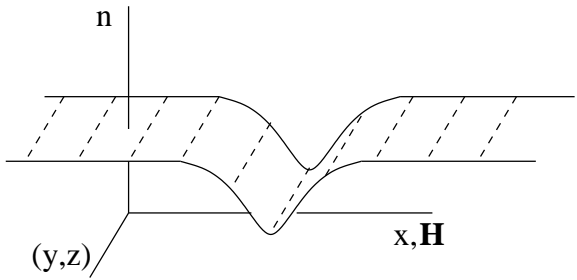}
\vspace*{0.1cm}
\caption{Schematic illustration of the density profile due to the soliton:
electron density depends only on the coordinate $x$ in the direction of the
magnetic field $H$; transverse to the field direction is labeled $(y,z)$.  The
phase profile (not shown) is a smeared step function (\ref{phase}); the step 
position coincides with that of the density dip.}
\end{figure} 
In dimensionless variables $\nu
= n/n_{0}$, $y = 2\pi ^{2}l^{2}n_{0}x$, $\tau = (2\pi
^{2}l^{2}n_{0})^{2}\hbar t/m$, $\zeta = y - \beta \tau$, the density ($\nu$) and phase ($\theta$)
profiles for the soliton are given by \cite{KNSQ}
\begin{equation}
\label{density}
\nu = 1-{3(1-\beta ^{2}) \over 2+(1+3\beta ^{2})^{1/2}\cosh [2(1- \beta ^{2})^{1/2}\zeta]}
\end{equation}
\begin{equation}
\label{phase}
2\theta = \cos^{-1}\left[{(3\beta ^{2}/\nu) - 1 \over (1 + 3\beta
^{2})^{1/2}}\right]
\end{equation}
where the dimensionless velocity $\beta$ is measured in units of the
sound velocity.  These formulas imply that as $\beta$ varies between
zero and unity, the soliton becomes more shallow, and the phase shift
across the soliton $\Delta \theta = \cos^{-1}[(3\beta ^{2} - 1)/(1 +
3\beta ^{2})^{1/2}]$ changes from $-\pi$ ($\beta = 0$) to zero ($\beta
= 1$); the soliton solution ceases to exist for velocities exceeding
the sound velocity.  Microscopically the domain wall is composed of
aligned point-like solitons belonging to different Landau tubes, and
there is a positive energy per unit area associated with the planar
defect.  It can be calculated as the energy of the soliton belonging
to the Landau tube (computed in \cite{KNSQ}), divided by the
cross-sectional area of the Landau tube $2\pi l^{2}$:
\begin{equation}
\label{senergy}
\sigma(\beta) = {\sqrt3(\pi l \hbar n_{0})^{2} \over m}(1-\beta^ {2})
\ln \left[{2+[3(1-\beta ^{2}]^{1/2} \over (1+3\beta ^{2})^{1/2}}\right]
\end{equation}
The positivity of the surface energy (\ref{senergy}) guarantees the
stability of the domain wall (\ref{density}), (\ref{phase}) against
transverse fluctuations.

{\it Stationary states:} - Solitons form a natural language to describe weakly 
excited states of the system.  For example, let us look at electron gas placed
inside a layer of macroscopic thickness $L$;  the plane of the layer is 
perpendicular to the direction of the magnetic field.  If the electrons cannot
leave the layer, then in the ground state the density profile will look as 
shown schematically in Figure 2 (left): the density stays constant almost 
everywhere except for the narrow vicinity of the walls where it sharply 
plunges to zero.
\begin{figure}[htbp]
\epsfxsize=3.0in
\vspace*{-0.3cm}
\hspace*{0.2cm}
\epsfbox{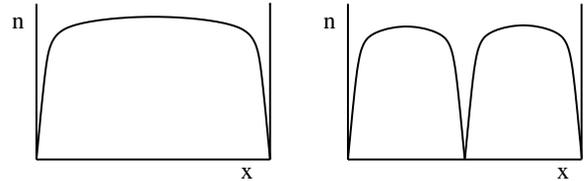}
\vspace*{0.1cm}
\caption{Sketch of the first (left) and the second (right) excited states of
electron gas in a box.}
\end{figure} 
It is convenient to think of these two dips as having one dark soliton present
in the system.  It is also convenient to take the convention that what is
shown in Figure 2 (left) is a first excited state with respect to the strictly 
uniform state which we will call the ground state.  Thus the first excited 
state has just one soliton present in the system.  Then the second excited 
state will have two solitons [Figure 2 (right)].  The argument can be
extended to any number of solitons.  For example, if there are {Q}
dark solitons present, then as long as soliton tails do not overlap,
$Q \ll l^{2}n_{0}L$, the system energy is given by $E(Q) = E_{0} +
S\sigma (0)Q$ where $E_{0}$ is the ground-state energy
(\ref{grstate}).  A similar conclusion is applicable to the case of a
ring-shaped sample with the only constraint that now $Q$ must be {\it even}
because the phase shift due to the dark soliton is $\pm \pi$.  Having an
even number of dark solitons will guarantee that the overall change of phase of the order parameter upon going around the ring is 
multiple of $2\pi$, and thus the order parameter is unique. 

Because translational symmetry is not broken for the ring geometry, dark
soliton states do not exhaust all possible excited states of the system: gray
solitons can also be involved.  However it will be easier to anticipate their
role after discussing the persistent current states.

{\it Persistent currents:} -  Persistent currents are only possible in the ring
geometry. Experimentally the magnetic field directed along the circumference 
of a ring-shaped sample can be generated by an electric current flowing 
along the symmetry axis perpendicular to the plane of the ring.  In the ring 
geometry the phase of the order parameter can only 
change by a multiple of $2\pi$.  The simplest way to accommodate a nonzero 
phase
change is to have a constant phase gradient along the ring \cite{Sonin}.  This
will lead to a spatially uniform persistent current state in which all the 
electrons are moving around the ring with the same velocity.  This state can be
also viewed as a stationary state.

There is also a novel nonuniform way of accommodating the phase difference.  
The corresponding
persistent current states can be constructed from gray solitons, and
they have a series of remarkable properties.  For example, imagine we
want to insert into the ring $Q$ solitons to accommodate an overall
phase shift of $2\pi$.  Then the phase shift due to the individual
soliton must be $2\pi/Q$.  This will in turn determine the soliton
velocity $\beta$ which will then fix the surface energy
(\ref{senergy}).  An example of four gray solitons moving along the
ring is sketched in Figure 3.  
\begin{figure}[htbp]
\epsfxsize=3.0in
\vspace*{-0.3cm}
\hspace*{0.2cm}
\epsfbox{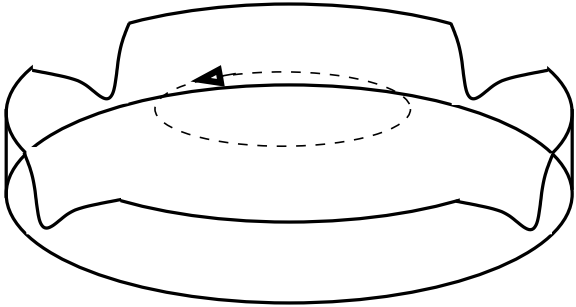}
\vspace*{0.1cm}
\caption{Schematic illustration of four solitons traveling on a ring.  The 
upper curve is the density profile, and the lower curve indicates the 
zero-density level.  The arrow shows the direction of motion of the soliton 
train.}
\end{figure} 
The energy of the train consisting of
$Q$ solitons running around the ring of circumference $L$ is given by
$E(Q) = E_{0} + S\sigma [\beta (Q)]Q$ as long as soliton tails do not
overlap, $Q \ll (1-\beta ^{2})^{1/2}l^{2}n_{0}L$.  In the reference
frame moving with the soliton train the solitons will be standing but
the background electron gas will be moving with a constant velocity.
This gives an example of a stationary state involving gray solitons.

For the uniform moving state the total angular momentum of
the system is quantized in units of $N\hbar$ where $N$ is the total
particle number
\cite{Sonin}.  We speculate that for the soliton train moving state the total 
angular momentum might be a {\it fraction} of $N\hbar$. Our contention is 
based on the
observation that while the soliton is moving, the electron motion is
localized only in the vicinity of the soliton core - merely a tiny
fraction of particles are involved in the motion.

In applying these results to the real system one has to keep in mind
that since a soliton is a density depletion, for solitons which do not
travel fast enough, the density at the soliton core can drop below the
left bound of (\ref{interval}).  Therefore our results are strictly
applicable only to solitons whose velocity is larger than some
critical velocity.  Future work is necessary to understand how
interactions between the electrons belonging to different Landau tubes
affect the slow solitons.  One possibility is that they convert the
depleted soliton core into a two-dimensional Wigner crystal.  Another
possibility is a modulational instability of the soliton in which
Coulomb repulsion of the electrons decreases at the expense of increase of the 
surface energy (\ref{senergy}).

The unique mathematical structure of the underlying equations 
(\ref{functional})-(\ref{eiler}) which so far did not enter the discussion
deserves a special commentary.  First we note that in the hydrodynamical limit 
when the last term of (\ref{eiler}) can be neglected (and $V = 0$) the 
problem of arbitrary motion of a one-dimensional gas with the equation of 
state $p \sim n^{3}$ can be solved in a closed form \cite{LL3}.  Second,  
(even for $V \sim x^{2}$) equations (\ref{nlse})-(\ref{eiler}) have 
self-similar solutions \cite{Rybin} which only exist for the equation of 
state $p \sim n^{3}$.  Physically some of them describe free 
expansion of the electron gas initially confined along the field direction 
\cite{KNSQ1}.

The author thanks T. J. Newman for valuable assistance, and 
X. Qi for useful discussions.  This work was supported by the 
Thomas F. Jeffress and Kate Miller Jeffress Memorial Trust.


\begin{references}

\bibitem{Ruder} 
Fields in the range $10^{9}$ to $10^{13} G$ exist in the vicinity
of white dwarfs and neutron stars; see H. Ruder, G. Wunner, H. Herold,
and F. Geyer, ``Atoms in Strong Magnetic Fields'', Springer-Verlag
Berlin Heidelberg, 1994, and references therein.

\bibitem{Lieb} 
E. H. Lieb, J. P. Solovej, and J. Yngvason, Phys. Rev. Lett. 69, 749
(1992), and references therein.

\bibitem{AM}
N. W. Ashcroft and N. D. Mermin, ``Solid State Physics'' (Saunders
College Publishing, 1976), Chapters 15 and 28.

\bibitem{Silin}
A. P. Silin, in ``Electron-Hole Droplets in Semiconductors'', Chapter 8,
p. 619, edited by C.D.Jeffries and L.V.Keldysh, North Holland, 1983,
Volume 6 of ``Modern Problems in Condensed Matter Sciences''.

\bibitem{LL1}
L. D. Landau and E. M. Lifshitz, ``Quantum Mechanics'', Volume 3, third
edition, Pergamon Press, 1977, Section 112.

\bibitem{McB}
A. H. MacDonald and G. W. Bryant, Phys. Rev. Lett. 58, 515 (1987), and
references therein.

\bibitem{Yak}
V. M. Yakovenko, Phys. Rev. B 47, 8851 (1993).

\bibitem{Abr}
A. A. Abrikosov, J. Low Temp. Phys. 2, 37 (1970); 2, 175 (1970).

\bibitem{Maslov}
C. Biagini, D. L. Maslov, M. Yu. Reizer, and L. I. Glazman, cond-mat/
0006407 v2.

\bibitem{KNSQ1}
E. B. Kolomeisky, T. J. Newman, J. P. Straley and X. Qi, in 
preparation.

\bibitem{Onis}
T. A. Onischenko, Trudy FIAN, 123, 7 (1980), L.V.Keldysh and T. A.
Onischenko, Pis'ma Zh. Eksp. Teor. Fiz. 24, 70 (1976) [Sov. Phys. JETP
Lett.  24, 59 (1976)].
 
\bibitem{Brack}
M. Brack and R. K. Bhaduri, ``Semiclassical Physics'' (Addison-Wesley
Publishing Company, Inc., Reading, Massachusetts, 1997), Chapter 4.

\bibitem{LL2}
L. D. Landau and E. M. Lifshitz, ``Fluid Mechanics'', (Pergamon, Oxford,
1987).

\bibitem{LP}
E. M. Lifshitz and L. P. Pitaevskii, ``Statistical Physics'', Part 2 (Pergamon,
Oxford, 1980).

\bibitem{KNSQ}
E. B. Kolomeisky, T. J. Newman, J. P. Straley and X. Qi, Phys. Rev. 
Lett. 85, 1146 (2000).

\bibitem{Kivshar}
Y. S. Kivshar and B. Luther-Davis, Phys. Rep. 298, 81 (1998).
  
\bibitem{Sonin}
E. B. Sonin, Zh. Eksp. Teor. Fiz. 59, 1416 (1970); F. Bloch, Phys. 
Rev. B 7, 2187 (1973).

\bibitem{LL3}
Section 105 of \cite{LL2}.

\bibitem{Rybin}
A. Rybin et al, cond-mat/0001059.

\end{references}
\end{document}